\documentclass[12pt]{article}
\usepackage{amstex,latexsym}

\numberwithin{equation}{section}

 \newcommand{\seta}{\longrightarrow}
 \newcommand{\eq}{\begin{equation}}
\newcommand{\real}{{\Bbb R}} \newcommand{\cpx}{{\Bbb C}}
\newcommand{\su}{{\frak s}{\frak u}(2)} \newcommand{\slc}{{\frak s}{\frak l}(2,\cpx)}
\newcommand{\so}{{\frak s}{\frak o}(2,1)} \newcommand{\SU}{SU(2)} 
\newcommand{\SO}{SO(2,1)}

\begin{document} \baselineskip16pt

\title{Finite action solutions of $SO(2,1)$ Hitchin's equations}
\author{Marcos Jardim \\ University of Pennsylvania \\ Department of Mathematics \\ 
209 South 33rd St. \\ Philadelphia, PA 19104-6395 USA}


\maketitle

\begin{abstract}
We present a 1-parameter family of finite action solutions to the
$S0(2,1)$ Hitchin's equations and explore some of its basic
properties. For a fixed value of the parameter, the solution is
smooth. We conclude by showing a multi-particle generalization of our basic solutions.
\end{abstract}

\pagebreak

\section{Introduction}
This brief paper is dedicated to find a special class of explicit solutions of the self-dual
Yang-Mills equations. These are originally defined on euclidean 4-space; the physically relevant
solutions are the ones with finite action and are called instantons. The same equations may
be dimensionally reduced to euclidean 3-space by imposing invariance under translations in one
direction. Such equations are also physically relevant, for its finite action solutions are
interpreted as magnetic monopoles. If we take one step further and consider the solutions
which are invariant under translations along two directions we obtain a set of equations in the
plane with no clear physical meaning.

Indeed, it is conjectured that there are no finite action solutions whatsoever to these
equations, with gauge group $SU(2)$ (see \cite{H}). For instance, if one tries to apply the
so-called 't Hooft ansatz \cite{JNR}, which produces the basic solutions to the higher
dimensional instanton and monopole equations, one will soon run into trouble. Recall that the
't Hooft ansatz reduces the self-dual Yang-Mills equation to find a nowhere vanishing
solution to the Laplace's equation. In dimensions 4 and 3, the fundamental solution of the
laplacian are proportional to
$\frac{1}{r^2}$ and $\frac{1}{r}$, respectively; but in dimension 2, such solution is
logarithmic, what makes the ansatz useless.

Nonetheless, these equations were extensively studied by Hitchin \cite{H} for gauge group $\SU$,
leading to the discovery of very interesting mathematical structures; similar results were
generalized to gauge groups $SU(n)$ by other authors.

We now take a different path, and show the existence of finite action solutions to the
so-called Hitchin's equations with gauge group $\SO$. Although there is no clear physical
meaning attached to these solutions, we expect that the solutions here presented might
inspire the study of gauge thoeries with non-compact gauge groups, as well as the search for less obvious physical interpretations.

\paragraph{Hitchin's equations.}
We begin with a brief review of Hitchin's equations.
Let $\real^4$ be parametrised by coordinates $(x,y;u,v)$. Consider a $SO(2,1)$ bundle $E\seta
\real^4$ with a connection, whose entries are $\so$ matrices depending only on the two first
coordinates:
\eq \label{c1}
A=A_1(x,y)dx+A_2(x,y)dy+\phi_1(x,y)du+\phi_2(x,y)dv
\end{equation}
Now let $\Phi=\phi_1+i\phi_2$ and $dz=du-idv$, hence:
\eq \label{c2}
A=A_1(x,y)dx+A_2(x,y)dy+\Phi(x,y)dz+\Phi^*(x,y)d\overline{z}
\end{equation}
Let $\widetilde{A}=A_1dx+A_2dy$; the self-duality equations are then given by:
\eq \label{heqs} \left\{ \begin{array}{l}
F_{\widetilde{A}}+[\Phi,\Phi^*]=0 \\
\overline{\partial}_{\widetilde{A}}\Phi=0
\end{array} \right. \end{equation}
which under this form are known as {\em Hitchin's equations}; these are now
regarded as equations on a two-dimensional plane. Here, $F_{\widetilde{A}}$ denotes
the curvature component along the $(x,y)$ plane. Conformal invariance of the self-duality
equations implies that such equations also make sense over any Riemann surface.

\section{$SO(2,1)$ solutions}
First, let us recall some of the elementary properties of the group $\SO$ and its Lie algebra
$\so$. Consider the usual Pauli matrices:
\eq \label{pauli} \begin{array}{ccc}
\sigma_1=\frac{1}{2}\left( \begin{array}{cc} i & 0 \\ 0 & -i \\  \end{array} \right) &
\sigma_2=\frac{1}{2}\left( \begin{array}{cc} 0 & 1 \\ -1 & 0 \\  \end{array} \right) &
\sigma_3=\frac{1}{2}\left( \begin{array}{cc} 0 & i \\ i & 0 \\  \end{array} \right)
\end{array} \end{equation}
which represent the generators of $\su$ in $End(\cpx^2)$ and satisfy the following
commutation relations:
\begin{equation} \left\{ \begin{array}{l}
\left[\sigma_1,\sigma_2\right]=\sigma_3 \\ \left[\sigma_1,\sigma_3\right]=-\sigma_2 \\
\left[\sigma_2,\sigma_3\right]=\sigma_1
\end{array} \right. \end{equation}

The Lie algebra $\so$ is defined by slightly different relations:
\begin{equation} \label{socommut} \left\{ \begin{array}{l}
\left[\tau_1,\tau_2\right]=\tau_3 \\ \left[\tau_1,\tau_3\right]=-\tau_2 \\
\left[\tau_2,\tau_3\right]=-\tau_1
\end{array} \right. \end{equation}
and can be represented in $End(\cpx^2)$ as:
\eq \label{so} \begin{array}{ccc}
\tau_1=\sigma_1 & \tau_2=i\sigma_2 & \tau_3=i\sigma_3
\end{array} \end{equation}
Note that choosing this representation is equivalent to regard the bundle $E\seta\real^4$
as a rank $2$ complex bundle.

Finally, recall that both $\{\sigma_1,\sigma_2,\sigma_3\}$ and $\{\tau_1,\tau_2,\tau_3\}$
generate the complex Lie algebra $\slc$; $\su$ is its unique compact real sub-algebra and
$\so$ is one of its non-compact real form. Hence, one can think of $\so$ as a 3-dimensional
plane sitting {\em diagonally} in the 6-dimensional space $\slc=\su\oplus i\su=\so\oplus i\so$.

As a group, $\SO$ is clearly not compact; its maximal compact
subgroup is the 1-parameter subgroup generated by $\tau_1$. $\SO$
is therefore homotopically equivalent to $S^1$, and has classifying
space $B\SO=\cpx{\Bbb P}^\infty$.

\paragraph{Non-compactness issues.}
One of the problems of working with a non-compact group is that the Killing form of its algebra
is not negative definite. In fact, using the commutation relations (\ref{socommut}) one quickly
verifies that the Killing form of $\SO$ is given by:
\eq \label{Kform}\begin{array}{ccc}
<\tau_i,\tau_j>=\eta_{ij}=\frac{1}{2}{\rm diag}(++-)
\end{array} \end{equation}
In particular, this implies that the action:
\eq \label{action}
{\cal S}(A)=-\int\left<F_A\wedge*F_A\right>
\end{equation}
is also indefinite, but {\em it is gauge-invariant}. Nonetheless, the solutions here presented
will be shown to have strictly positive action density, when computed with the natural pairing
above.

An alternative approach would be to note that although $\SO$ is
non-compact, its algebra has a negative definite bilinear form, and
we might use such form to compute the action (\ref{action}). The
pairing we have in mind is:
\eq \label{altKform}
<\tau_i,\tau_j>={\rm Tr}(\tau_i\overline{\tau_j})=-\frac{1}{2}\delta_{ij}
\end{equation}
where by $\overline{\tau_j}$ we mean the complex conjugate to $\tau_j$. Now, the expression
(\ref{action}) is always strictly positive.
The problem is that this pairing has no invariant meaning and depends on a choice of basis
on $\cpx^2$. Hence, it might happen that a gauge transformation of a connection with finite
action result in a connection with divergent action.

In the next paragraph, we present our 1-parameter family of solutions to (\ref{heqs}) with gauge group
$\SO$ and then we proceed to compute its action with respect the Killing form (\ref{Kform}).

\paragraph{The ansatz.}
Our starting point is the following ansatz; reparametrize $(x,y)$, the first two coordinates of
$\real^4$, by polar coordinates $(r,\theta)$ and consider the connection:
\eq \label{ansatz}
A=f(r).\tau_1d\theta+g(r).\tau_2du+h(r).\tau_3dv
\end{equation}
Note that such ansatz is more general than it seems, for most connections can be put in this
form after gauge transformations. First, gauge away the $dr$ component; then, if the three
remaining components are linearly independent, apply constant gauge transformations so that they
lie along the generators of the algebra.

In terms of the fields involved on Hitchin's formulation, we have that the ansatz:
\eq \label{hitansatz} \begin{array}{ccc}
\widetilde{A}=-yf(x,y).\tau_1dx+xf(x,y).\tau_1dy & \ \ &
\Phi= g(x,y).\tau_2+ih(x,y).\tau_3
\end{array} \end{equation}

The equations (\ref{heqs}) are then given by:
\eq \label{anseqs} \left\{ \begin{array}{l}
\frac{1}{r}\frac{df}{dr}-gh=0 \\
\frac{dg}{dr}+\frac{1}{r}fh=0 \\
\frac{dh}{dr}+\frac{1}{r}fg=0
\end{array} \right. \end{equation}

To integrate this system, suppose that $g=h$, change coordinates to $r=e^{-t}$ and let
$F=1-f$ and $G=e^{-t}g$. Then (\ref{anseqs}) are reduced to:
\eq \label{anseqs2} \left\{ \begin{array}{l}
\frac{dF}{dt}-G^2=0 \\
\frac{dG}{dt}+FG=0 \\
\end{array} \right. \end{equation}
whose solutions are
\eq \label{soln} \begin{array}{ccc}
F(t)=c\tanh(ct) & \ & G(t)=c\ {\rm sech}(ct)
\end{array} \end{equation}
for any constant $c>0$. Changing these back to the $r$ coordinate, get:
\eq \label{soln2} \begin{array}{ccc}
f(r)=\frac{(1-c)-(1+c)r^{2c}}{1+r^{2c}} & \ & g(r)=h(r)=2c\frac{r^{c-1}}{1+r^{2c}}
\end{array} \end{equation}
and substituting these in (\ref{ansatz}) we have:
\begin{equation} \label{soln3}
A = \frac{(1-c)-(1+c)r^{2c}}{1+r^{2c}}\tau_1d\theta+2c\frac{r^{c-1}}{1+r^{2c}}
(\tau_2du+\tau_3dv)
\end{equation}
which is the promised 1-parameter family of solutions to the $\SO$ Hitchin's equations.

Note that this is not the unique solution to the system (\ref{heqs}). One could, for instance,
set $g=-h$ after the same change of coordinates, but setting $F=f+1$, obtain:
\eq \label{anseqs3} \left\{ \begin{array}{l}
\frac{dF}{dt}+G^2=0 \\
\frac{dG}{dt}+FG=0 \\
\end{array} \right. \end{equation}
whose solutions are:
\eq \label{solnx} \begin{array}{ccc}
F(t)=c\coth(ct) & \ & G(t)=c\ {\rm csch}(ct)
\end{array} \end{equation}
for any $c>0$. In terms of the original coordinates, we get:
\eq \label{solnx2} \begin{array}{ccc}
f(r)=\frac{(c-1)+(c+1)r^{2c}}{1-r^{2c}} & \ & g(r)=-h(r)=2c\frac{r^{c-1}}{1-r^{2c}}
\end{array} \end{equation}
but the singularity at $r=1$ makes this solution useless for our purposes.

\ref{soln2}\section{Some properties.}
We analyze some of the properties of the solutions (\ref{soln2}). First, we show that they have
finite action. Then, we observe that for $c=1$ the solution (\ref{soln2}) is smooth
but singular for any other value of $c$. Finally, we compute its holonomy around an
arbitrarily large disc centered at the origin and show multi-particle generalizations of our
smooth solution.

\paragraph{Computing the action.}
We use the Killing form (\ref{Kform}) to compute (\ref{action}) and show that although
the bilinear form (\ref{Kform}) is not positive definite, the action of (\ref{soln3}) has
positive density. Indeed, plugging (\ref{soln3}) into (\ref{action}), we have:
\eq \label{calcaction}
{\cal S}(A)=\int_{\real^2}\frac{1}{r^2}\left(\frac{df}{dr}\right)^2dxdy=
2\pi c^4\int_0^\infty\frac{r^{4c}}{r^3(1+r^{2c})^4}dr
\end{equation}
which is clearly convergent for any $\frac{1}{2}<c<\infty$. In particular, if $c=1$, we have:
\eq \label{c1action}
{\cal S}(A)=2\pi\int_0^\infty\frac{r}{(1+r^{2})^4}dr=\frac{\pi}{3}
\end{equation}
As we pointed out before, the above quantity has invarinat meaning and is indepent of the
choice of gauge. Note also that all connections of the form (\ref{ansatz}) has strictly
positive action.

\paragraph{Smooth and singular solutions.}
In Euclidean coordinates, the solution (\ref{soln3}) can be written as:
\begin{eqnarray} A & = &  
\frac{(1-c)-(1+c)(x^2+y^2)^c}{1+(x^2+y^2)^c}\cdot\frac{-y\tau_1dx+x\tau_1dy}{x^2+y^2}
+ \nonumber \\
& & \ \ + \frac{c(x^2+y^2)^{\frac{c-1}{2}}}{1+(x^2+y^2)^c}\cdot(\tau_2du+\tau_3dv)
\label{soln4} \end{eqnarray}
and one can see that $A$ is smooth if and only if $c=1$. Writing it explicitly, the
smooth, finite energy solution of Hitchin's equations that motivated the present paper:
\begin{equation} \label{soln5} 
\widetilde{A} = \frac{2}{1+x^2+y^2}(-ydx+xdy)\tau_1  \ \ \ \
\Phi = \frac{1}{1+x^2+y^2}(\tau_2+i\tau_3)
\end{equation}

For other values of the parameter, such that $c>1$, then $A$ has a singularity of codimension 2
at $x=y=0$ of type $\frac{1}{r}$. Such singular solutions were studied by several authors in
dimensions two and four (see, for instance, \cite{KM} and \cite{FHP}), and the interested reader
should refer to these works. This type of singular field configuration is also known as
{\em meron}, for they carry fractional topological charge.

\paragraph{Limiting holonomy.}
As mentioned before, we want to compute the holonomy of (\ref{soln3}) around an arbitrarily
large circle centered at the origin. More precisely, we want to solve the initial value
problem parametrized by the radial distance $r$ for $\gamma_r:S^1\seta\SO$:
\eq \label{ivp} \left\{ \begin{array}{l}
\frac{d}{d\theta}\gamma_r+A_\theta\gamma_r=0 \\ \gamma(0)=I
\end{array} \right. \end{equation}

Using expression (\ref{soln2}), it is easy to see that:
\eq  \label{lim} \lim_{r\seta\infty}\gamma_r(\theta)=
\left( \begin{array}{cc} \exp\left(\frac{1}{2}i(1+c)\theta\right) & 0
\\ 0 & \exp\left(-\frac{1}{2}i(1+c)\theta\right)
\end{array} \right) \end{equation}

Such procedure also allows us to characterize multi-particle solutions, just like the usual
multi-instanton solutions on $\real^4$ are characterized through the degree of a mapping from
the 3-sphere at infinity to $SU(2)\equiv S^3$. The difference in the present case is that
$\gamma_\infty(\theta)$ represents a map from the circle at infinity  of the plane to
$S^1\subset\SO$, regarded as the maximal compact sub-group of $\SO$, which, as we have mentioned
before, classifies $\SO$-bundles topologically.

\paragraph{Multi-instanton solutions.}
Again, fix $c=1$. The smooth solution (\ref{soln5}) might be generalized as follows:
\begin{eqnarray}  A & = & 
\sum_{k=1}^N\frac{2}{1+(x-x^k)^2+(y-y_k)^2}\cdot 
\frac{-(y-y_k)\tau_1dx+(x-x_k)\tau_1dy}{(x-x_k)^2+(y-y_k)^2} + \nonumber \\
& & \frac{2}{1+(x-x_k)^2+(y-y_k)^2}(\tau_2du+\tau_3dv) \label{soln6} 
\end{eqnarray}
The $\gamma_\infty(\theta)$ map (\ref{lim}) associated to this solution is then given by:
\begin{equation} \label{lim2} \left( \begin{array}{cc}
e^{iN\theta} & 0 \\ 0 & e^{-iN\theta}
\end{array} \right) \end{equation}
whose degree is $N$. Following the above analogy, we can interpret solutions of the form
(\ref{soln6}) as a {\em multi-particle solution}. Each point $(x_k,y_k)\in\real^2$ corresponds
to the position of a {\em particle}.

It hence easy to conclude that the space of solutions of
the $\SO$ Hitchin's equations (\ref{heqs}), up to gauge equivalence, is at least
a $2N$-dimensional manifold, parametrized by the coordinates $(x_k,y_k)$.

A still more general non-smooth, multi-particle solution can be obtained by superposing
instantons with different values of the parameter $c$:
\begin{eqnarray}  A & = & 
\sum_{k=1}^N\frac{(1-c_k)-(1+c_k)((x-x_k)^2+(y-y_k)^2)^c_k}{1+((x-x_k)^2+(y-y_k)^2)^c_k}
\cdot \nonumber \\ & & \ \ \ \ \ \ 
\cdot\frac{-(y-y_k)\tau_1dx+(x-x_k)\tau_1dy}{(x-x_k)^2+(y-y_k)^2} + \nonumber \\
& & \ \ \ \ \ \ + 
2c_k\frac{((x-x_k)^2+(y-y_k)^2)^{\frac{c_k-1}{2}}}{1+((x-x_k)^2+(y-y_k)^2)^c_k}
(\tau_2du+\tau_3dv) \label{soln7}
\end{eqnarray}
The interpretation of the $\gamma_\infty(\theta)$ map (\ref{lim}) now is less obvious, but
it may be understood as counting fractionally charged particles.

\paragraph{Conclusion.} We have shown that although there are no known finite action $SU(2)$
solutions of Hitchin's equations, it is possible to write down finite action solutions of the
$\SO$ version of these equations. It seems likely that the non-compactness of the structural
group has a deeper role. Such issue is certainly worth of further investigation. For instance,
are there finite action $SU(n)$ solutions, for $n\geq2$? What about the non-compact forms
$SO(p,n-p)$?

Another point would be to determine the correct framework which gives a natural interpretations
to the equations and solutions presented above. One might also expect to find some relation
with the well-known phenomenon of fractional statistics on $\real^2$.

On the mathematical side, it is interesting to ask how the Chern-Weil theory adapts to
non-compact gauge groups, what would fit in the bigger programme of understanding the gauge
theory of non-compact Lie groups. The $SO(2,1)$ seems a good choice for acquiring some intuition
on such an unexplored subject.

\smallskip

\paragraph{Acknowledgments.} This work was supported by CNPq, Brazil.

\bibliographystyle{plain}  \end{document}